\documentclass[acmsmall]{acmart}
\AtBeginDocument{%
  }

\usepackage[graphicx]{realboxes}
\usepackage{color}
\usepackage{tcolorbox}
\usepackage{xcolor}
\usepackage{bm}
\usepackage{multirow}
\usepackage{ulem}
\usepackage{threeparttable}
\usepackage{array}
\usepackage{xspace}
\usepackage{mdframed}
\usepackage{makecell}
\newcounter{finding}

\usepackage{amsmath}
\usepackage{graphicx}
\usepackage{float}
\usepackage{subfigure}
\usepackage{booktabs}
\usepackage{pifont}

\usepackage[linesnumbered,ruled,vlined]{algorithm2e}

\newcommand{\finding}[1]{\refstepcounter{finding}
 \begin{mdframed}[linecolor=gray,roundcorner=12pt,backgroundcolor=gray!15,linewidth=3pt,innerleftmargin=2pt, skipabove=10pt, skipbelow=10pt, leftmargin=0cm,rightmargin=0cm,topline=false,bottomline=false,rightline = false]
  \textbf{Finding.} #1
 \end{mdframed}
}

\setcopyright{none}
\begin{document}

\title{Do Advanced Language Models Eliminate the Need for Prompt Engineering in Software Engineering?}

\author{Guoqing Wang}
\email{guoqingwang@stu.pku.edu.cn}
\affiliation{%
  \institution{Key Lab of HCST (PKU), MOE; \\ 
  School of Computer Science, Peking University}
  \city{Beijing}
  \country{China}
}

\author{Zeyu Sun}
\email{zeyu.zys@gmail.com}
\affiliation{%
 \institution{National Key Laboratory of Space Integrated Information System, Institute of Software, Chinese Academy of Sciences}
  \city{Beijing}
  \country{China}
}

\author{Zhihao Gong}
\email{zhihaogong@stu.pku.edu.cn}
\affiliation{%
  \institution{Key Lab of HCST (PKU), MOE; \\ 
  School of Computer Science, Peking University}
  \city{Beijing}
  \country{China}
}

\author{Sixiang Ye}
\affiliation{%
  \institution{Beijing University of Chemical Technology}
  \city{Beijing}
  \country{China}}
\email{yesx.sxye@gmail.com}

\author{Yizhou Chen}
\email{yizhouchen@stu.pku.edu.cn}
\affiliation{%
  \institution{Key Lab of HCST (PKU), MOE; \\ 
  School of Computer Science, Peking University}
  \city{Beijing}
  \country{China}
}

\author{Yifan Zhao}
\email{zhaoyifan@stu.pku.edu.cn}
\affiliation{%
  \institution{Key Lab of HCST (PKU), MOE; \\ 
  School of Computer Science, Peking University}
  \city{Beijing}
  \country{China}
}

\author{Qingyuan Liang}
\email{liangqy@stu.pku.edu.cn}
\affiliation{%
  \institution{Key Lab of HCST (PKU), MOE; \\ 
  School of Computer Science, Peking University}
  \city{Beijing}
  \country{China}
}

\author{Dan Hao}
\email{haodan@pku.edu.cn}
\affiliation{%
  \institution{Key Lab of HCST (PKU), MOE; \\ 
  School of Computer Science, Peking University}
  \city{Beijing}
  \country{China}
}

\renewcommand{\shortauthors}{Wang et al.}

\begin{abstract}
Large Language Models (LLMs) have significantly advanced software engineering (SE) tasks, with prompt engineering techniques enhancing their performance in code-related areas. However, the rapid development of foundational LLMs such as the non-reasoning model GPT-4o and the reasoning model o1 raises questions about the continued effectiveness of these prompt engineering techniques. This paper presents an extensive empirical study that reevaluates various prompt engineering techniques within the context of these advanced LLMs.
Focusing on three representative SE tasks, i.e., code generation, code translation, and code summarization, we assess whether prompt engineering techniques still yield improvements with advanced models, the actual effectiveness of reasoning models compared to non-reasoning models, and whether the benefits of using these advanced models justify their increased costs. Our findings reveal that prompt engineering techniques developed for earlier LLMs may provide diminished benefits or even hinder performance when applied to advanced models. In reasoning LLMs, the ability of sophisticated built-in reasoning reduces the impact of complex prompts, sometimes making simple zero-shot prompting more effective.
Furthermore, while reasoning models outperform non-reasoning models in tasks requiring complex reasoning, they offer minimal advantages in tasks that do not need reasoning and may incur unnecessary costs. Based on our study, we provide practical guidance for practitioners on selecting appropriate prompt engineering techniques and foundational LLMs, considering factors such as task requirements, operational costs, and environmental impact. Our work contributes to a deeper understanding of effectively harnessing advanced LLMs in SE tasks, informing future research and application development.

\end{abstract}


\begin{CCSXML}
<ccs2012>
   <concept>
       <concept_id>10011007.10011074.10011099.10011102.10011103</concept_id>
       <concept_desc>Software and its engineering~Software testing and debugging</concept_desc>
       <concept_significance>500</concept_significance>
       </concept>
   <concept>
       <concept_id>10011007.10011074.10011099.10011102</concept_id>
       <concept_desc>Software and its engineering~Software defect analysis</concept_desc>
       <concept_significance>500</concept_significance>
       </concept>
 </ccs2012>
\end{CCSXML}

\ccsdesc[500]{Software and its engineering~Software testing and debugging}
\ccsdesc[500]{Software and its engineering~Software defect analysis}

\keywords{Large Language Model}

\received{20 February 2007}
\received[revised]{12 March 2009}
\received[accepted]{5 June 2009}

\maketitle

\section{Introduction}

Large Language Models (LLMs)~\cite{chatgpt, deepseekv2, claude, gemini, llama} have achieved remarkable results across various domains~\cite{wu2024survey, yan2024llm-evaluator, qin2023large, satpute2024can, chen2024deep}, demonstrating human-like intelligence, especially in natural language processing (NLP)~\cite{sun2023sentiment, yi2024survey, mo2024large}. This success has prompted a growing number of Software Engineering (SE) researchers to integrate LLMs into solving diverse SE tasks, yielding promising outcomes~\cite{sun2024source,xia2023keep, chen2024deep, liu2024large, dong2024self, kang2023explainable, kang2024quantitative}. Despite these successes, significant challenges persist in effectively using LLMs for task completion~\cite{satpute2024can, yu2024fight, koo2023benchmarking, chen2024deep}. 

Harnessing LLMs in SE to maximize their exceptional in-context learning and reasoning capabilities~\cite{gao2023makes,yao2022react, wei2022chain, wang2023element}, relies on the prompts used, the information provided, and the specific ways in which models are invoked~\cite{sun2024source, xu2023expertprompting, zhong2024debug}. In response, various strategies have emerged, often termed ``prompt engineering techniques'', which aim to optimize LLM performance beyond simple model calls. These techniques\footnote{To avoid confusion, following previous work~\cite{sun2024source}, the term ``techniques'' in this paper specifically refers to prompt engineering techniques, while specific approaches based on these techniques will be referred to as ``approaches''.} include few-shot prompting~\cite{gao2023makes}, Chain-of-Thought (CoT) prompting~\cite{wei2022chain, yao2022react}, critique prompting~\cite{ma2024combining, kim2024language}, expert prompting~\cite{xu2023expertprompting}, and so on. While these techniques have proven effective across different software development and maintenance tasks, such as code generation, code understanding, software testing, and debugging, challenges like hallucinations and inaccuracies remain~\cite{zhong2024debug, dong2024self, pan2024lost, chen2024deep}.
To address these limitations, researchers have shifted focus to dynamic strategies~\cite{dong2024self, huang2023agentcoder, pan2024lost, yang2024exploring} involving continuous interaction with the LLMs, task decomposition, and result verification. Building upon these concepts, recent approaches employ prompt engineering techniques such as multi-agent systems~\cite{ma2024combining, dong2024self}, iterative refinement processes~\cite{ma2024combining, yang2024exploring, zhong2024debug}, and the integration of additional contextual information to refine the LLM's output~\cite{zhong2024debug, yang2024exploring}. By leveraging these prompt engineering techniques, LLMs can deliver more reliable results in complex SE tasks, paving the way for further advancements in SE research.

With the rapid advancements in large language model training~\cite{chatgpt, deepseekv2, claude, gemini, llama}, foundational models are being iterated and updated at an accelerated pace~\cite{gpt4o, o1, o1-mini}. More advanced foundational models demonstrate improved understanding and generating capabilities. When OpenAI released the GPT-4o model~\cite{gpt4o}, its performance outperformed that of most prompt engineering techniques developed for earlier foundational LLMs in coding tasks~\cite{paperswithcode_humaneval_code_generation}. The subsequent o1, o1-mini models~\cite{o1,o1-mini} integrate CoT reasoning, allowing reasoning-based LLMs to autonomously decompose complex problems into a series of simpler steps, thereby forming effective strategies for tackling intricate logical issues. 
However, many prompt engineering techniques for code~\cite{gao2023makes, xu2023expertprompting, dong2024self} were developed based on the capabilities of the earlier model, ChatGPT-3.5~\cite{chatgpt}, as it was the only option available at the time. This overlooks the enhancements offered by the more advanced GPT-4o~\cite{gpt4o} and the reasoning capabilities of the o1 and o1-mini models~\cite{o1,o1-mini}. Moreover, OpenAI's guidance indicates that using complex prompts is not recommended for reasoning LLMs~\cite{o1, o1-mini}. Thus, this raises the first question about \ding{172} {\em the effectiveness of these prompt engineering techniques on the more advanced models}. Furthermore, while it is claimed that the reasoning LLMs, i.e., o1 and o1-mini, may provide enhanced performance, \ding{173} {\em what is its actual effectiveness compared to non-reasoning models, and what are its respective advantages and disadvantages}? Additionally, the reasoning LLMs typically incur higher operational costs, both in terms of monetary expenditure and time efficiency~\cite{o1, openai_pricing}. In addition to computational and token-based costs, the varying levels of carbon emissions are a critical consideration.
This raises the third question: \ding{174} {\em do the benefits of utilizing these advanced models justify their increased costs?}

To explore these questions, this paper presents the first extensive study aimed at revisiting and re-evaluating a variety of prompt engineering techniques within the context of the GPT-4o and o1-mini\footnote{Due to the high cost of the o1-preview~\cite{o1,openai_pricing}, we utilize the o1-mini~\cite{o1-mini} as a replacement.} models. 
\textbf{For Question~\ding{172}}, we investigate the effectiveness of these prompt engineering techniques on more advanced models, assessing whether they still yield significant improvements or if methodological adjustments are required for adaptation to new LLMs. 
\textbf{For Question~\ding{173}}, we examine the actual performance of the reasoning LLM, compared to non-reasoning models, exploring its advantages and disadvantages, particularly in light of its capability to autonomously decompose complex problems and form reasoning strategies.
\textbf{For Question~\ding{174}}, we provide practical guidance to users on selecting appropriate prompt engineering techniques and foundational LLMs based on our findings, particularly considering whether the benefits of utilizing these advanced models justify their increased operational costs.

To comprehensively assess the effectiveness of prompt engineering strategies within advanced LLMs, we deliberately select three representative code-related tasks~\cite{chen2024deep, liu2024large} based on their prevalence and significance in the SE field: code generations~\cite{zhong2024debug,huang2023agentcoder,dong2024self}, code translation~\cite{ahmad2023summarize,pan2024lost,yang2024exploring}, and code summarization~\cite{gao2023makes,xu2023expertprompting,sun2024source,ahmed2024automatic, wang2023element,kim2024language}. These tasks correspond to the common scenarios of text-to-code, code-to-code, and code-to-text in SE research~\cite{codexglue}, respectively, and they encompass a wide range of practical applications.
To evaluate these approaches, we used three widely recognized datasets: HumanEval~\cite{chen2021codex} for code generation, CodeTrans~\cite{yang2024exploring} for code translation, and CodeSearchNet~\cite{husain2019codesearchnet} of CodeXGLUE~\cite{codexglue} benchmark for code summarization. These datasets are standard benchmarks in the field and provide a solid foundation for comparative analysis.
For each task, we identify and include several state-of-the-art approaches that utilize distinct prompt engineering techniques, ensuring a diverse and representative evaluation. Specifically, we select 11 approaches~\cite{zhong2024debug,huang2023agentcoder,dong2024self, ahmad2023summarize,pan2024lost,yang2024exploring,gao2023makes,xu2023expertprompting,sun2024source,ahmed2024automatic, wang2023element,kim2024language} that employ techniques such as few-shot prompting, CoT prompting, critique prompting, and multi-agent collaboration. 
We replace the underlying LLMs with the latest foundational models, including both non-reasoning and reasoning models~\cite{gpt4o,o1-mini}, to evaluate their performance across these SE tasks. This comprehensive assessment aims to provide new insights into the effectiveness of prompt engineering strategies within advanced LLMs.

From this study, we uncover several notable findings, some of which are summarized as follows:

\begin{enumerate}
    \item {\bf Effectiveness of Prompt Engineering Techniques:} When applied to more advanced LLMs, the improvements achieved by prompt engineering techniques developed for earlier LLMs are often less pronounced than those reported in prior studies. In some cases, these techniques even introduce performance drawbacks. For the reasoning LLM, the LLM’s ability to self-correct through internal reasoning makes most prompt engineering techniques less effective than using a simple zero-shot prompt. The formulation of prompts itself is less critical than effectively leveraging accurate information and addressing errors based on reliable feedback, such as execution details in software testing. 
    \item { \bf The actual effectiveness of reasoning models compared to non-reasoning models:} For tasks that need many steps of reasoning, the reasoning LLMs can achieve better performance than non-reasoning LLMs. For tasks that do not require complex reasoning, performance differences between reasoning models and non-reasoning models are minimal, even worse than prompt engineering techniques based on non-reasoning LLMs. The output format and content of reasoning LLMs are more flexible and longer, which may be unnecessary and hard to handle.
    \item { \bf Practical guidance to users on utilizing these advanced models: } Given the additional costs, time efficiency,  and potential environmental impacts associated with reasoning models, non-reasoning models may be a more cost-effective option for tasks where advanced reasoning capabilities are not essential. When the expected output is not long, such as code summarization, it is recommended to use non-reasoning LLMs. Furthermore, when utilizing reasoning LLMs, the expected format and content should be restricted precisely.
\end{enumerate}

In summary, this paper makes the following contributions:

\begin{itemize} 
    \item To the best of our knowledge, this study is the first to empirically evaluate a diverse range of prompt engineering techniques on more advanced LLMs within code-related tasks, specifically focusing on code generation, code translation, and code summarization. It includes an evaluation of the non-reasoning and reasoning LLMs.
    \item We provide a detailed analysis of the performance of approaches leveraging various prompt engineering techniques. We find that some prompt engineering techniques will decrease and lose their effectiveness on advanced LLMs, and the reasoning LLMs can not always outperform non-reasoning LLMs.
    \item Based on our findings, we offer insights and implications that can guide the adoption and future development of LLMs in SE tasks. We also provide practical guidance to users when selecting prompt engineering techniques and foundational LLMs, considering the monetary expenditure, time costs, and potential impact on the environment. 
\end{itemize}

\section{Background and Related Work}

This section reviews the prompt engineering techniques~\cite{gao2023makes, xu2023expertprompting,wei2022chain,kim2024language,dong2024self} used in selected code-related tasks~\cite{zhong2024debug,huang2023agentcoder,dong2024self, ahmad2023summarize,pan2024lost,yang2024exploring,gao2023makes,xu2023expertprompting,sun2024source,ahmed2024automatic, wang2023element,kim2024language} in our work, i.e., code generation, code translation, and code summarization.

The most basic approach is zero-shot prompting, where LLMs receive simple instructions to perform a task~\cite{sun2024source}, such as:
\textit{``Please generate a short comment for the following function: <code>.''}
Earlier LLMs often struggled with this kind of simple prompts~\cite{gao2023makes,xu2023expertprompting} due to limited capabilities, resulting in unsatisfactory performance. To enhance these models, researchers introduce more sophisticated prompting techniques that improve the models' performance in SE tasks, such as few-shot prompting~\cite{gao2023makes,pan2024lost,huang2023agentcoder,ahmad2023summarize}, CoT prompting~\cite{wei2022chain,yao2022react,wang2023element}, and expert prompting~\cite{xu2023expertprompting}.

Few-shot prompting, also known as in-context learning~\cite{gao2023makes}, involves providing models with a few examples in addition to the task instructions~\cite{gao2023makes,pan2024lost,huang2023agentcoder,ahmad2023summarize}. These examples serve as conditioning inputs, enabling the model to better understand tasks and generate more accurate responses. Few-shot prompting has proven effective across many code-related tasks~\cite{pan2024lost,huang2023agentcoder,ahmad2023summarize,yang2024exploring}. Previous work~\cite{gao2023makes} studies what kind of examples to provide, how many examples to provide, and in what order can deliver better effects.
For tasks requiring complex reasoning, researchers develop CoT prompting. CoT guides LLMs through intermediate reasoning steps~\cite{wei2022chain,yao2022react}, breaking tasks into logical segments. This process allows the model to emulate human-like reasoning, improving its ability to handle more complicated code-related problems~\cite{wang2023element}. 
Prompt formulation has a significant impact on model performance, as noted in prior studies~\cite{xu2023expertprompting,gao2023makes,huang2023agentcoder}. This led to the development of expert prompting, which begins with a description of an ``expert'', such as \textit{``You are an expert on software programming $\ldots$'' }. This description is then integrated with the task and problem content, aligning the model’s output with expert-level understanding and improving task performance~\cite{xu2023expertprompting,sun2024source}.

In many code-related tasks, generating correct outputs often requires more than a single model call, as the initial responses may contain errors~\cite{dong2024self,zhong2024debug}. Critique prompting addresses this issue by prompting LLMs to identify and correct errors in their responses~\cite{kim2024language}. This iterative process may involve multiple rounds of corrections or coordination between agents.
Furthermore, some tasks benefit from the inclusion of domain knowledge. For instance, execution information can assist in fixing bugs in code generation and code translation~\cite{yang2024exploring,pan2024lost,zhong2024debug,huang2023agentcoder}. Incorporating task-specific information enhances the accuracy and reliability of LLMs in code-related tasks.
Hence, many recent LLM-based frameworks integrate multiple prompting strategies, such as AgentCoder~\cite{huang2023agentcoder} and LDB~\cite{zhong2024debug} in code generation, and Unitrans~\cite{yang2024exploring} and LostinTransl~\cite{pan2024lost} in code translation.

\section{Study Design}

\subsection{Research Questions}
This study aims to address the following research questions:

\textbf{RQ1: Are previous prompt engineering techniques effective for more advanced LLMs?}
Existing studies on LLM-based code-related tasks in Software Engineering have predominantly utilized earlier models like ChatGPT-3.5 and GPT-4, which are now outdated and no longer supported by OpenAI~\cite{chatgpt}. Recent evaluations~\cite{paperswithcode_humaneval_code_generation} have shown that approaches based on these earlier LLMs sometimes underperform compared to more advanced LLMs that are utilized without specific prompt engineering techniques. Advanced models, trained with higher-quality data and more robust training strategies, may exhibit reduced sensitivity to prompt engineering techniques. Notably, the o1 and o1-mini models incorporate CoT reasoning~\cite{o1,o1-mini}, enabling them to autonomously decompose complex problems into simpler steps and devise reasoning strategies for solving intricate logical tasks~\cite{o1}. This raises the question of whether prompt engineering techniques remain necessary to enhance performance in such advanced models. Thus, this research question aims to identify which types of prompt engineering techniques, if any, continue to be effective with more advanced LLMs (including non-reasoning LLMs and reasoning LLMs).  

\textbf{RQ2: What are the advantages and limitations of the reasoning LLMs in SE tasks compared to previous LLMs?}
Building on the results from RQ1, this question seeks to explore the specific conditions under which the reasoning model outperforms non-reasoning LLMs in SE tasks. Given the diversity of SE tasks and scenarios, it is essential to determine the types of data and tasks where the reasoning models excel and where they may encounter difficulties. By investigating these factors, we aim to provide a comprehensive understanding of the reasoning models' strengths and weaknesses in the SE domain.

\textbf{RQ3: How should practitioners and researchers select prompt engineering techniques and foundational LLMs, considering the monetary expenditure and time costs?}
OpenAI's pricing indicates that reasoning LLMs have significantly higher costs and longer API response times compared to non-reasoning LLMs~\cite{openai_pricing}. Moreover, due to the model's autonomous reasoning phase, which users cannot directly control, accurately estimating the total cost of using reasoning LLMs becomes challenging. Therefore, this research question aims to evaluate the balance between cost and effectiveness in SE tasks, providing recommendations on when and how the reasoning LLMs should be used in practice. We aim to offer practical guidance for practitioners and researchers to make informed decisions regarding the adoption of advanced LLMs in cost-sensitive contexts.

% task dataset
\subsection{Experimental Setup}

Since LLMs in Software Engineering are primarily employed in code-related tasks~\cite{liu2024large,chen2024deep}, which can be categorized into three scenarios based on previous research~\cite{codexglue}, i.e., text-to-code, code-to-code, and code-to-text. Hence, we evaluate prompt engineering techniques across these three scenarios. Specifically, we select one representative task for each scenario~\cite{liu2024large,chen2024deep, codexglue, husain2019codesearchnet, huang2023agentcoder, dong2024self, ahmad2023summarize, pan2024lost, yang2024exploring, gao2023makes, xu2023expertprompting, sun2024source, ahmed2024automatic, wang2023element, kim2024language}: code generation for text-to-code, code translation for code-to-code, and code summarization code-to-text, respectively. For each task, we select several state-of-the-art approaches based on different prompt engineering techniques. Due to the high cost of utilizing LLMs, we can not include all approaches.
Our selection principle for approaches is to only consider approaches that are published, and we limit the timeline to 2023 and beyond. Additionally, the approaches should provide a reproducible package, and the approaches can run normally when replacing the underlying LLMs with the latest foundational LLMs. 
Considering that many advanced approaches on the public leaderboard for code generation have not been published~\cite{paperswithcode_humaneval_code_generation}, we selected the top-ranked approach~\cite{huang2023agentcoder} among them for comparison.
As mentioned before, many recent LLM-based approaches integrate multiple prompt engineering techniques. To further clarify these selected approaches~\cite{zhong2024debug,huang2023agentcoder,dong2024self, ahmad2023summarize,pan2024lost,yang2024exploring,gao2023makes,xu2023expertprompting,sun2024source,ahmed2024automatic, wang2023element,kim2024language}, we provide a summary table~\ref{tab:technique_summary} that outlines the prompt engineering techniques utilized by each approach. 

For each task, we provide a detailed description of the selected state-of-the-art approaches, the datasets employed, and the evaluation metrics in the following part. 

\begin{table*}[t]
    \caption{The summary of the prompt engineering techniques utilized by each approach.}
    \label{tab:technique_summary}
    \resizebox{0.9\textwidth}{!}{%
    \begin{tabular}{llcccccccc}
    \toprule
   &   & Few-shot & CoT & Critique & Expert & Multi-agent & Multi-iteration & Domain knowledge  \\ \midrule
\multicolumn{1}{l}{Code Generation}&\multicolumn{1}{l|}{CoT~\cite{wei2022chain}} && $\star$ &&&&&&   \\ 
&\multicolumn{1}{l|}{AgentCoder~\cite{huang2023agentcoder}}  & $\star$ &&&& $\star$ & $\star$ & $\star$  \\ 
&\multicolumn{1}{l|}{Self-collaboration~\cite{dong2024self}}  &&&&& $\star$ & $\star$ & $\star$   \\ 
&\multicolumn{1}{l|}{LDB~\cite{zhong2024debug}}  &&&&& $\star$ & $\star$ & $\star$   \\ \midrule
\multicolumn{1}{l}{Code Translation}&\multicolumn{1}{l|}{S\&G~\cite{ahmad2023summarize}} && $\star$ &&&&& \\ 
&\multicolumn{1}{l|}{Unitrans~\cite{yang2024exploring}}  & $\star$  &&&& $\star$ & $\star$ & $\star$   \\ 
&\multicolumn{1}{l|}{LostinTransl~\cite{pan2024lost}}  &&&&&& $\star$ & $\star$  \\ \midrule
\multicolumn{1}{l}{Code Summarization}&\multicolumn{1}{l|}{Few-shot~\cite{gao2023makes}} & $\star$ &&&&&&  \\
&\multicolumn{1}{l|}{CoT~\cite{wang2023element}} && $\star$ &&&&& \\
&\multicolumn{1}{l|}{Critique~\cite{kim2024language,sun2024source}} &&& $\star$ &&&& \\
&\multicolumn{1}{l|}{Expert~\cite{xu2023expertprompting}}  &&&& $\star$ &&& \\
&\multicolumn{1}{l|}{ASAP~\cite{ahmed2024automatic}} & $\star$ &&&&&& $\star$ \\
\bottomrule
    \end{tabular}
    }
\end{table*}

\subsubsection{Code Generation}
\noindent

\paragraph{A. Approaches}
In our experiments, we compare four state-of-the-art code generation approaches: Chain-of-Thought (CoT)~\cite{wei2022chain}, AgentCoder~\cite{huang2023agentcoder}, Self-collaboration~\cite{dong2024self}, and LDB~\cite{zhong2024debug}. 

\textit{Chain-of-Thought.} We utilize CoT prompt engineering technique based on a standard CoT prompt reported by previous work~\cite{wei2022chain,sun2024source}. For the problem description, the LLM is asked to first generate the chain of thought for the problem. Subsequently, the chain of thought is attached to the description of the code to be generated as the input prompt.

\textit{AgentCoder.} AgentCoder is a multi-agent framework designed for code generation that incorporates specialized agents, including a test designer, programmer, and test executor. The test designer agent generates test cases for the code produced, while the test executor runs these test cases and provides feedback to the programmer agent. This iterative process allows for continuous code refinement based on the feedback loop, enhancing the quality of the generated code.

\textit{Self-collaboration.} Self-collaboration employs three distinct LLM agents, i.e., an analyst, a coder, and a tester, each acting as a specialized ``expert'' in software development tasks, covering analysis, coding, and testing stages, respectively. This framework facilitates collaboration among virtual agents, forming a team that works together autonomously to tackle code generation tasks, reducing the need for human intervention.

\textit{LDB.} LDB is an LLM-based framework that refines generated programs by incorporating runtime execution information. It segments code into basic blocks and tracks intermediate variable values after each block during execution. This approach enables the model to focus on smaller units of code, verify correctness against the task description incrementally, and efficiently identify errors throughout the code execution flow.

\paragraph{B. Dataset}
We conduct our evaluation on the HumanEval benchmark~\cite{chen2021codex}, a widely-used dataset for assessing the code generation capabilities of LLMs. HumanEval consists of 164 programming problems, each defined by a natural language prompt and paired with a reference solution. The dataset includes unit tests to verify the correctness of the generated code, making it suitable for evaluating various code generation approaches based on different prompt engineering techniques.  

\paragraph{C. Metric}
The primary metric for evaluating model performance is pass@k, which measures the proportion of problems solved correctly within $k$ attempts~\cite{chen2021codex}. This metric emphasizes the model's ability to generate syntactically and semantically accurate Python code. For our evaluation, we specifically use pass@1, which calculates the success rate based on the model’s top-1 prediction, aligning with recent studies~\cite{zhong2024debug,huang2023agentcoder,dong2024self,paperswithcode_humaneval_code_generation}. This metric provides a clear measure of the model's precision in generating accurate solutions within a single attempt, reflecting real-world use cases where efficiency and correctness in the initial generation are critical.

\subsubsection{Code Translation}
\noindent

\paragraph{A. Approaches}
In our evaluation of the code translation task, we compare three state-of-the-art approaches: Summarize-and-Generate (S\&G)~\cite{ahmad2023summarize}, Unitrans~\cite{yang2024exploring}, and LostinTransl~\cite{pan2024lost}.

\textit{Summarize-and-Generate (S\&G).}
Summarize-and-Generate (S\&G) is initially proposed by Ahmad et al.\cite{ahmad2023summarize} as a strategy for enhancing unsupervised program translation. In this paper, we adopt the S\&G paradigm by prompting large language models (LLMs) to first generate summaries of the source code and subsequently produce the translated code as a kind of CoT prompting. This two-step process leverages the summarization capabilities of LLMs to capture the semantic essence of the original program, which we hypothesize may improve the quality of the generated translations.

\textit{Unitrans.}
UniTrans~\cite{yang2024exploring} first crafts a series of test cases for target programs with the assistance of source programs. It then leverages these auto-generated test cases to augment the code translation process and evaluate their correctness through execution. Subsequently, UniTrans iteratively repairs incorrectly translated programs based on the outcomes of test case executions.

\textit{LostinTransl.}
Similar to UniTrans, LostinTransl\cite{pan2024lost} introduces an iterative repair strategy for code translation, hypothesizing that incorporating additional contextual information in prompts can enhance translations generated by LLMs. To achieve this, they include elements such as incorrect translations, error details, and expected behavior. Unlike UniTrans, LostinTransl assesses translation accuracy and obtains execution feedback directly from the original dataset's test cases.

\paragraph{B. Dataset}
We use a refined version of the dataset initially released by Roziere et al.\cite{roziere2020unsupervised}, which comprises 948 parallel code functions in C++, Java, and Python, sourced from the GeeksforGeeks platform. Yang et al.\cite{yang2024exploring} identified notable errors and inconsistencies within the original dataset and subsequently conducted an extensive data-cleaning process, resulting in a curated version with 568 parallel code samples. To enhance the reliability of our evaluation, we adopt this refined dataset released by Yang et al.\cite{yang2024exploring}. Considering the high computational cost of evaluation, we focus on the code translation between two popular programming languages, i.e., Java and Python.

\paragraph{C. Metric}
Following previous studies\cite{yang2024exploring,zhu2024semi}, we adopt the execution-based metric, Computational Accuracy (CA) for evaluation. CA measures the proportion of successfully passed test cases, providing a direct measure of the functional correctness of translated programs. We exclude static metrics such as BLEU \cite{bleu,bleu_norm} and CodeBLEU \cite{ren2020codebleu}, as they focus on surface-level syntactic similarity and do not effectively capture the semantic equivalence essential for accurate code translation.

\subsubsection{Code summarization}
\noindent

\paragraph{A. Approaches}
We compare five widely used prompt engineering techniques in code summarization: few-shot, Chain-of-Thought, critique, expert, and ASAP~\cite{gao2023makes,wang2023element, sun2024source,kim2024language,xu2023expertprompting,ahmed2024automatic}. 

\textit{Few-Shot.} The few-shot prompting is proposed by Gao et al.~\cite{gao2023makes}, and is specifically designed for code summarization. The example selection principle and order of the example\footnote{The number of examples is set to 4 as their findings.} remain the same as their implementation.

\textit{Chain-of-Thought.} We utilize the CoT steps in code summarization as previous work does~\cite{wang2023element, sun2024source}. The technique first asks LLMs to answer five questions (i.e., the name of the function, the input parameters that are being accepted by the function, the expected output or return value of the function, the specific requirements or constraints for using this function, and the additional dependencies or external requirements). Based on the response, the technique integrates the above information and asks LLMs to generate comments.

\textit{Critique.} The critique prompting~\cite{kim2024language} improves the quality of LLMs’ answers by asking LLMs to find errors in the previous answers and correct them. 

\textit{Expert.} The expert prompting~\cite{xu2023expertprompting, sun2024source} first asks LLMs to generate a description of an expert who can complete the instruction, and then the description serves as the system prompt for zero-shot prompting. To generate the description of an expert, the prompt engineering technique employs few-shot prompting to let LLMs generate a description of an expert who can ``Generate a short comment in one sentence for a function''.

\textit{ASAP.} ASAP~\cite{ahmed2024automatic} leverages multiple prompt engineering techniques. It employs few-shot prompting to identify relevant examples based on BM25~\cite{robertson2009probabilistic}. Subsequently, it extracts semantic features for each code sample (including the target function and the retrieved exemplars), including the repository name, the fully qualified name of the target function, its signature, the Abstract Syntax Tree (AST) tags of its identifiers, and its data flow graph to enhance the generation process.

\paragraph{B. Dataset}
We perform our evaluation on the CodeSearchNet dataset~\cite{husain2019codesearchnet} of CodeXGLUE~\cite{codexglue}, a widely-used benchmark in many code-related tasks~\cite{lin2023cct5, zhu2024grammart5,chen2024deep,liu2024large}, including code summarization~\cite{gao2023makes,sun2024source,ahmed2024automatic}. The code summarization dataset encompasses six programming languages and contains a test set of approximately 53,000 samples. Similar to code translation, to mitigate the high computational cost of evaluation, we focus on two popular programming languages, i.e., Java and Python, and randomly select 250 samples from each language for our experiments.

\paragraph{C. Metric}
Previous research~\cite{sun2024source,ahmed2024automatic,eliseeva2023commit,wanghard} has shown that common automated evaluation methods for code summarization, such as those based on summary-summary text similarity or semantic similarity, often lack consistency with human evaluations. In contrast, GPT-based evaluation methods have demonstrated a stronger correlation with human judgments.
As a result, we adopt the GPT-based evaluation method, following the approach of Sun et al.~\cite{sun2024source}, which provides evaluation code and metrics that better align with human assessments. The LLM rates each summary from 1 to 5 where a higher score represents a higher quality of the summary.
This method allows us to more accurately measure the quality of LLM-generated summaries by comparing them to human-like reference summaries, ensuring a more reliable evaluation of model performance.

Finally, to assess the fundamental capabilities of LLMs, we serve zero-shot prompting as a baseline for all three tasks, which utilizes models without any specific prompting strategies. The zero-shot prompt for code generation is defined as the description in HumanEval~\cite{chen2021codex}, which is the same as the previous work's setting~\cite{dong2024self,huang2023agentcoder,zhong2024debug}. For code translation, the zero-shot prompt is the same as the prompt provided in the previous evaluation of Transcoder by Yang et al.~\cite{yang2024exploring}. The zero-shot prompt for code summarization is provided by Sun et al.~\cite{sun2024source}.

\section{Results}

In this section, we present the evaluation results to answer each RQ. We implement the selected approaches based on the released code of their reproducible packages and only change the fundamental LLM to the more advanced non-reasoning and reasoning models. We select GPT-4o as the representation of non-reasoning models because of its excellent performance and affordable cost~\cite{gpt4o,openai_pricing}. For reasoning models, due to the high cost of the o1-preview~\cite{o1,openai_pricing}, we utilize the o1-mini~\cite{o1-mini} as a replacement. Regarding the randomness of the response of LLMs, we run each approach three times and present the average results for each experiment.

\subsection{RQ1: the effectiveness of each prompt engineering technique}
To answer RQ1 regarding the effectiveness of each prompt engineering technique utilized in code-related tasks, we present the code generation results in Table~\ref{tab:RQ1_gene}, the code translation results in Table~\ref{tab:RQ1_trans}, and the code summarization results in Table~\ref{tab:RQ1_summ}. 
In the code generation task, applying the CoT prompt engineering technique to o1-mini encounters a limitation due to OpenAI's constraints, which prevent the model from generating a chain of thought~\cite{chatgptweb,o1,o1-mini}. To address this, we incorporate the chain of thought generated by GPT-4o as part of the input for o1-mini. For the code translation task, LLMs may lack familiarity with the precise code format required after translation, causing the format of the translation does not meet the evaluation criteria. 
Although the output should ideally be a single function, translation often results in a full class structure instead, increasing the risk of format inconsistencies without a guiding example.
To mitigate this, we provide a single example (1-shot) to help guide LLMs in generating correctly formatted code. 

\begin{table*}[t]
    \caption{Performance comparison of different approaches in code generation.}
    \label{tab:RQ1_gene}
    \resizebox{0.4\textwidth}{!}{%
    \begin{tabular}{lcc}
    \toprule
       & GPT-4o & o1-mini  \\ \midrule
\multicolumn{1}{l|}{Zero-shot}  & 90.4      & 93.9  \\
\multicolumn{1}{l|}{CoT} & 91.5      & 94.1   \\ 
\multicolumn{1}{l|}{AgentCoder} &  96.3     &  95.1  \\ 
\multicolumn{1}{l|}{Self-collaboration} & 90.9      & 95.7   \\ 
\multicolumn{1}{l|}{LDB} & 94.5      &  96.3  \\ \midrule
\multicolumn{1}{l|}{AgentCoder-no-iter} &  87.8     &  89.6  \\ 
\multicolumn{1}{l|}{Self-collaboration-no-iter} & 90.2   & 94.1   \\
\multicolumn{1}{l|}{LDB-no-iter} & 91.5      & 94.4   \\ 

\bottomrule
    \end{tabular}
    }
\end{table*}

\begin{table*}[t]
    \caption{Performance comparison of different approaches in code translation.}
    \label{tab:RQ1_trans}
    \resizebox{0.55\textwidth}{!}{%
    \begin{tabular}{lcccc}
    \toprule
& \multicolumn{2}{c}{Java2Python} & \multicolumn{2}{c}{Python2Java}  \\ \midrule

       & GPT-4o             & o1-mini        & GPT-4o             & o1-mini        \\ \midrule
\multicolumn{1}{l|}{zero-shot}& 0.947       & 0.943       & 0              & 0.085        \\
\multicolumn{1}{l|}{1-shot} & 0.944       & 0.954       & 0.776       & 0.803       \\
\multicolumn{1}{l|}{S\&G} & 0.941       & 0.943       & 0.660       & 0.099       \\
\multicolumn{1}{l|}{Unitrans} & 0.949   & 0.908      & 0.817       & 0.682       \\
\multicolumn{1}{l|}{LostinTransl}  & 0.971  & 0.978   & 0.783   & 0.806  \\ \midrule
\multicolumn{1}{l|}{Unitrans-no-iter}  & 0.941   & 0.900    & 0.783   & 0.548        \\
\multicolumn{1}{l|}{LostinTransl-no-iter}  & 0.942 & 0.956    & 0.782  & 0.799        \\
\bottomrule
    \end{tabular}
    }
\end{table*}

\begin{table*}[t]
    \caption{Performance comparison of different approaches in code summarization.}
    \label{tab:RQ1_summ}
    \resizebox{0.45\textwidth}{!}{%
    \begin{tabular}{lcccc}
    \toprule
& \multicolumn{2}{c}{Java} & \multicolumn{2}{c}{Python}  \\ \midrule

       & GPT-4o             & o1-mini        & GPT-4o             & o1-mini        \\ \midrule
\multicolumn{1}{l|}{Zero-shot}  & 4.14       & 4.23       & 3.71       & 4.12        \\
\multicolumn{1}{l|}{Few-shot}   & 4.19       & 4.15       & 3.89       & 4.00       \\
\multicolumn{1}{l|}{CoT}  & 4.26       & 3.98       & 4.31       & 4.09       \\
\multicolumn{1}{l|}{Critique}    & 4.42       & 3.76       & 4.46       & 3.71        \\
\multicolumn{1}{l|}{Expert} & 4.44       & 3.98       & 4.26       & 4.04       \\
\multicolumn{1}{l|}{ASAP}   & 4.34       & 4.11       & 4.41       & 4.08        \\
 \bottomrule
    \end{tabular}
    }
\end{table*}

As shown in these tables, the performance of the tested approaches varies significantly between GPT-4o and o1-mini. Most prompt engineering techniques demonstrate effectiveness \textbf{when applied to GPT-4o, generally leading to performance improvements over the zero-shot baseline}. However, the magnitude of these gains is notably smaller compared to previous LLMs. In contrast, \textbf{only a few prompt engineering techniques yield positive results with o1-mini}, and many techniques either fail to produce improvements or result in performance drops.

In the code generation task, for GPT-4o, AgentCoder achieves a pass@1 rate of 96.3\%, while LDB reaches 94.5\%, suggesting that iterative and collaborative prompting strategies enhance performance. CoT prompting achieves a pass@1 rate of 91.5\%, showing only a modest improvement over the zero-shot baseline of 90.4\%. This indicates that CoT prompting may not be essential for code generation when using more advanced LLMs.
Similarly, in the code translation task, the performance of S\&G and Unitrans remains similar to that of the simple prompt when translating from Java to Python. When translating from Python to Java, S\&G performs even worse than the simple prompt, highlighting that advanced non-reasoning LLMs like GPT-4o can directly understand and translate code to the target language without extensive guidance. However, iterative and collaborative prompting methods like Unitrans and LostinTransl outperform the simple prompt, indicating that such strategies can still benefit code translation in non-reasoning LLMs.
In code summarization, all prompt engineering techniques outperform zero-shot prompting. Among these, critique prompting appears to be the most effective technique for GPT-4o, achieving scores of 4.42 and 4.46 for Java and Python, respectively.

For o1-mini, which has a built-in CoT strategy, the zero-shot baseline already achieves a high success rate of 93.9\% in code generation. 
Given that the CoT prompting performance is nearly identical to the zero-shot baseline, we conduct a Wilcoxon signed-rank test~\cite{wilcoxon1945individual}. The resulting p-value exceeds 0.10, indicating no significant difference between the two results.
However, other selected approaches, i.e., AgentCoder, Self-collaboration, and LDB, can still improve performance, although the extent of improvement becomes smaller compared to GPT-4o. We suspect that the included execution information contributes to the enhancement of the performance.
This result suggests that \textbf{providing explicit CoT prompts is unnecessary for o1-mini}, as its built-in CoT handles reasoning effectively.

Moreover, in the code translation task, prompt engineering techniques have a limited impact when utilizing o1-mini. Except for LostinTransl\footnote{Although LostinTransl and Unitrans are both iterative and collaborative approaches, LostinTransl iterates based on the execution feedback from the evaluated dataset's test cases.}, techniques like S\&G and Unitrans negatively affect performance compared to simple prompts. An even more pronounced trend is observed in code summarization, where the basic zero-shot prompt not only competes with but typically exceeds the results of more complex prompt engineering strategies when using o1-mini. This indicates that \textbf{in some contexts, especially in tasks utilizing reasoning LLMs like o1-mini, simpler approaches may be more effective} than more intricate prompt engineering techniques.

We also notice that while o1-mini demonstrates better performance in zero-shot scenarios compared to GPT-4o, when \textbf{enhanced with prompt engineering techniques, GPT-4o often exceeds o1-mini's performance}. For example, with the AgentCoder technique, GPT-4o reaches a pass@1 rate of 96.3\% (the highest pass@1 in code generation), whereas o1-mini achieves 95.1\%. This trend is even more pronounced in tasks such as Java code summarization, where several approaches on GPT-4o, including CoT, Critique, Expert, and ASAP, outperform all configurations on o1-mini. This suggests that the sophisticated built-in reasoning capabilities of o1-mini yield diminishing returns when further enhanced with prompt engineering, whereas GPT-4o, with a lower baseline, benefits more significantly from the same techniques, often outperforming o1-mini post-enhancement.

\finding{The results demonstrate that while prompt engineering can still enhance the performance of non-reasoning models like GPT-4o, the benefits are significantly reduced. For example, in code generation, GPT-4o's pass@1 rate increases modestly from 90.4\% (zero-shot) to 96.3\% with AgentCoder. In contrast, the reasoning model o1-mini achieves a high pass@1 rate of 93.9\% with zero-shot prompting, and prompt engineering techniques offer minimal or no improvement. This suggests that for reasoning models like o1-mini, prompt engineering may have diminishing returns or even negative impacts. Additionally, tasks like code summarization do not significantly benefit from using reasoning models.}

In the results of code generation and code translation, the approaches that can outperform zero-shot prompting on o1-mini are AgentCoder, Self-collaboration, LDB, and LostinTransl. 
We observe that although the formulation of prompts in the approaches are different from each other, they all utilize multi-iteration based on the execution information of test cases. Hence, we raise a question as to whether the execution information contributes to the enhancement of the performance compared to zero-shot prompting.
To verify our hypothesis, we remove the feedback of the test execution information and the iteration phase in code generation and code translation\footnote{There is no approach in code summarization utilizing execution information.}. The results are listed in Table~\ref{tab:RQ1_gene} and \ref{tab:RQ1_trans}, donated as ``<Approach>-no-iter''. 

As shown in these tables, for the non-reasoning model, GPT-4o, when removing the feedback of the test execution information and the iteration phase, the performance of each approach is similar to or lower than the performance of zero-shot prompting. It indicates that \textbf{the useful part of each approach's prompt is the test execution information and the fix phase during the iteration} instead of the formulation of prompts. 

For o1-mini, when removing the feedback of the test execution information and the iteration phase, the performance of each approach is also similar to or lower than the performance of simple promptings, such as zero-shot and 1-shot prompting. It shows that even without the complex description of prompts, LLMs with built-in CoT can understand the problem and reason the correct solution to the problem by itself. 

In code translation, we observe a significant performance decline for Unitrans when using o1-mini. Specifically, the performance drops from 0.954 to 0.900 when translating from Java to Python and from 0.682 to 0.548 for Python-to-Java. Unitrans initially generates a series of test cases for the target program and uses these cases to enhance code translation. However, after a manual check, we find that this approach can be error-prone, as the test generation phase may not accurately represent the original code, leading to incomplete or flawed test cases. This issue is amplified when correctness is not verified and errors are corrected only during execution, potentially causing o1-mini to reason inaccurately about the target code, resulting in incorrect translations. 

A similar challenge arises in code summarization. ASAP, which aims to improve summarization by extracting semantic features from each code sample, performs worse with o1-mini than with simple zero-shot prompting. This suggests that, without execution feedback that can reflect the ground truth, the supplemental information extracted from the input may not enhance performance and can even hinder it. For advanced LLMs like o1-mini, which possess sophisticated internal reasoning mechanisms, such information may be unnecessary for code summarization. Inputting all information simultaneously, without considering its relevance, can disrupt the internal Chain-of-Thought logic, leading to degraded performance.

\finding{Our results show that the specific wording of prompts has minimal impact on advanced models like GPT-4o and o1-mini. Performance gains are primarily due to real execution feedback used during iteration. For example, in code generation, removing execution feedback in AgentCoder reduces GPT-4o's pass@1 rate from 96.3\% to 87.8\%. Conversely, providing inaccurate information without actual execution feedback can mislead reasoning models and degrade performance.}

\subsection{RQ2: the advantages and limitations of the reasoning LLMs in SE tasks}
To address RQ2, we delve deeper into the performance of the reasoning LLMs. 
We aim to explain its effectiveness and highlight differences compared to user-designed prompt engineering techniques.
Additionally, recognizing that o1-mini does not perform optimally across all code-related tasks, we categorize the types of cases where the reasoning model tends to fail and summarize the underlying reasons for these failures. This analysis helps us identify the model's limitations and provides insights into areas where prompt engineering or model refinement may be necessary. 

\paragraph{Advantages}
We find that \textbf{for problems involving multiple steps of reasoning, LLMs with built-in CoT generally outperform non-reasoning LLMs}. To further verify our finding, we extract the steps to solve problems of o1-mini using OpenAI's Chat Website~\cite{chatgptweb} as the steps of CoT reasoning\footnote{Since the details of o1-mini's CoT can not be obtained, we can only use the abbreviated CoT displayed on the website as the object of our statistics.}. To be specific, for each task, we randomly choose 100 samples, a total of 300 samples. 
The average length of CoT's step is 3.52 in code generation, 4.35 in code translation, and 1.38 in code summarization. Hence, we suspect that code generation and code translation need more reasoning steps than code summarization for LLMs.
We inspect the differences in effectiveness between GPT-4o and o1-mini with different lengths of CoT. We filter the problems that the length of o1-mini CoT's step is longer than or equal to 5. We find that for these problems, the performance of o1-mini is 16.67\% better than GPT-4o. For the problems where the length of o1-mini CoT's step is shorter than 5, the performance of o1-mini is 2.89\% better than GPT-4o. It indicates that the non-reasoning models, like GPT-4o, cannot conduct complex reasoning, which causes the performance of o1-mini to be better than GPT-4o in code generation.

\begin{figure*}[t]
  \centering
  \includegraphics[width=\linewidth]{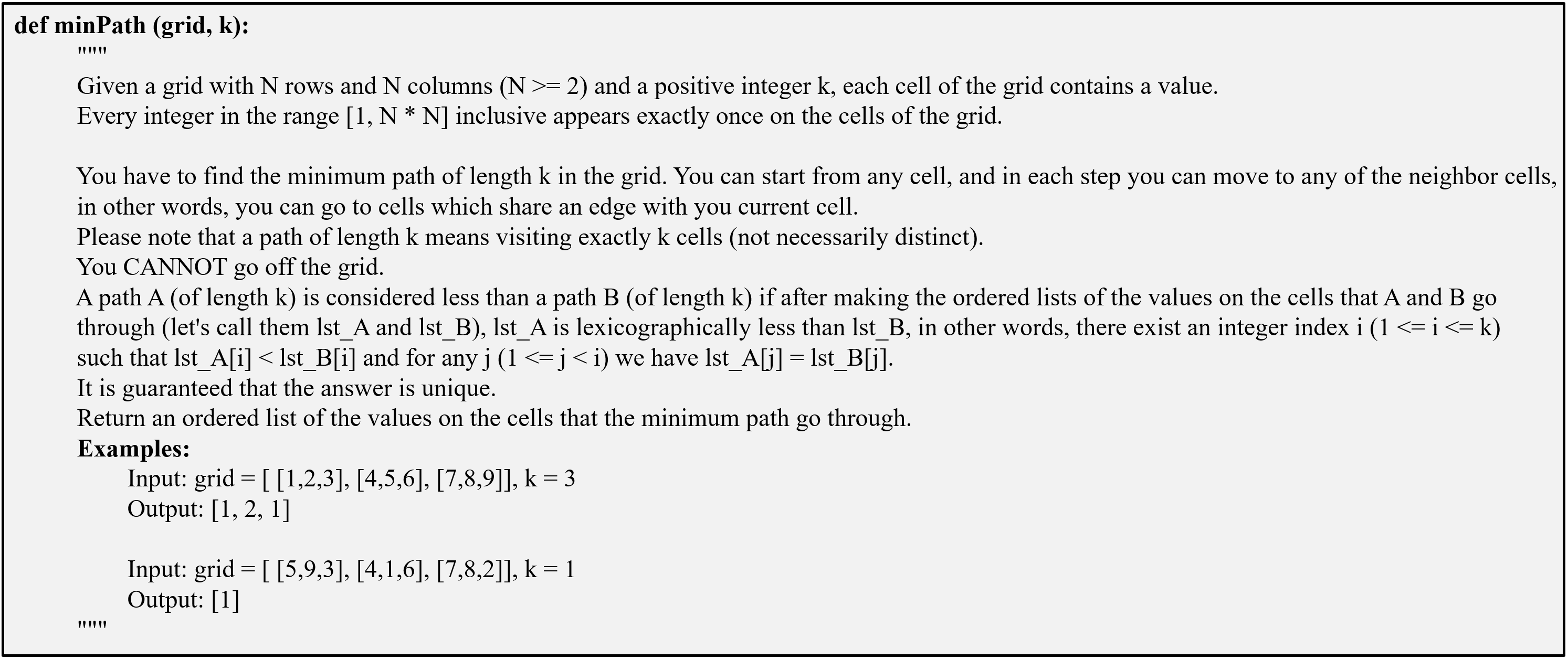}
  \caption{Problem 129 in HumanEval.}
  \label{case1}
\end{figure*}

\begin{figure*}[t]
  \centering
  \includegraphics[width=\linewidth]{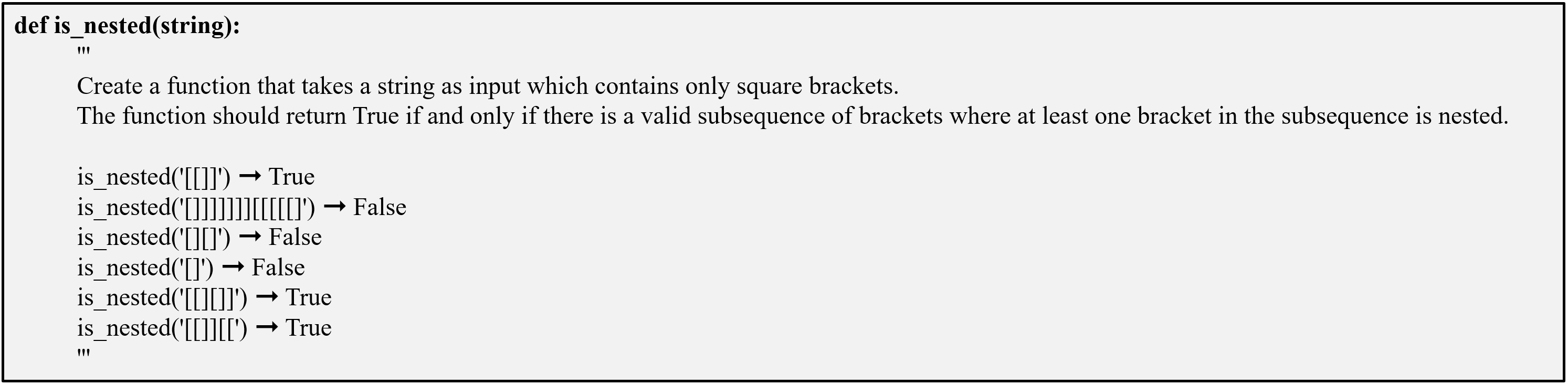}
  \caption{Problem 132 in HumanEval.}
  \label{case2}
\end{figure*}

An example of $HumanEval 129$ illustrates the difference between GPT-4o and o1-mini.
The problem asks to find a path of length $k$ in an $N \times N$ grid where $N \ge 2$ (see Fig.~\ref{case1} for details).  
For the problem, o1-mini solves the problem using a dynamic programming (DP) algorithm, effectively considering time complexity and the size of the search space through 10 reasoning steps in its CoT. In contrast, GPT-4o employs a depth-first search (DFS) algorithm that ultimately results in a timeout due to inefficiency. Even when advanced prompt engineering techniques like Self-collaboration and LDB are applied to GPT-4o, the solution still times out. This example demonstrates that reasoning LLMs like o1-mini can provide proper and efficient solutions based on the problem's difficulty and features—capabilities not present in GPT-4o without built-in CoT.

Another example of $HumanEval 132$ in Fig.~\ref{case2} demonstrates the influence of prompt engineering techniques on reasoning LLMs. The task requires writing a function is\_nested that takes a string of square brackets ($[]$) and determines whether it contains a valid nested pair. Only o1-mini with zero-shot prompting generates the correct answer. It takes o1-mini 32 reasoning steps over 35.17 seconds, during which its understanding deepens as it repeatedly verifies and refines its solution. If it identifies issues, it regenerates to correct them. However, when using prompt engineering techniques like AgentCoder and Self-collaboration, o1-mini is hindered by the complex prompts, leading to implementations that overlook certain special cases.

Analyzing the reasons behind these results, we find that \textbf{o1-mini's built-in CoT capability allows it to break down complex problems into manageable steps and correct errors iteratively}, particularly in scenarios with longer CoT sequences. This adaptability enables the model to match the depth of its reasoning to the complexity of the task. For problems that require extensive reasoning, o1-mini leverages its CoT more effectively than models without such capabilities, optimizing the reasoning process without overcomplicating simpler problems. This adaptability and built-in error correction make it less reliant on external prompt engineering to achieve high performance, especially in tasks that demand a deeper understanding and solution exploration.

\finding{ Our analysis shows that for problems requiring multi-step reasoning, specifically when the CoT length is 5 steps or more, reasoning LLMs like o1-mini outperform non-reasoning LLMs by an average of 16.67\%. 
Otherwise, the performance advantage decreases to 2.89\%. Further analysis shows that it probably due to the built-in CoT capability allows it to break down complex problems into manageable steps and correct errors iteratively. }

\paragraph{Limitations}
We further examine the limitations found in reasoning LLMs, specifically focusing on o1-mini. Despite its advanced capabilities, we find that \textbf{o1-mini often exhibits excessive divergent thinking}, leading to overextended reasoning in straightforward tasks. In our analysis, where the CoT length is less than three steps, we observe that in 24\% of cases where o1-mini underperformed compared to GPT-4o, the issue stems from unnecessary and expansive reasoning. This tendency to consider irrelevant factors complicates the model’s decision-making process and negatively impacts its effectiveness in simpler tasks.

Additionally, 
\textbf{o1-mini is less structured in its reply formats}. Across multiple randomized experiments, we identify a significant issue: with HumanEval's default prompt, o1-mini occasionally regenerates code segments that should remain unmodified. Specifically, nearly 40\% of o1-mini's incorrect answers under zero-shot prompting are deemed incorrect because their output formats can not be processed by standard post-processing tools used in previous evaluations, compared to 0\% for GPT-4o.
This indicates that o1-mini may modify parts of the input that are meant to be fixed, a problem not encountered with GPT-4o.

In code summarization tasks, when provided with the same prompts, o1-mini's responses consistently contain reasoning descriptions rather than directly providing concise answers, as GPT-4o does. This necessitates additional post-processing to extract the relevant comment parts from the given code, adding extra steps to the workflow.

Furthermore, when provided with human-written prompts, \textbf{reasoning models like o1-mini may need additional time to interpret these instructions, potentially causing errors.}
This mismatch between the model's internal reasoning process and external prompts can diminish the effectiveness of human-written prompts or even introduce negative impacts on performance. We suspect that this difference in thinking and understanding results in human-crafted prompts having limited or adverse effects on reasoning models. The autonomous ability of these models to think, correct, and rethink allows them to handle complex problems more effectively without the need for intricate prompts, highlighting the diminishing returns of traditional prompt engineering techniques when applied to advanced reasoning LLMs like o1-mini.

\finding{ 
o1-mini often engages in unnecessary and expansive reasoning, particularly in simpler tasks where the CoT is less extensive. In scenarios where the CoT length is under 3 steps, o1-mini underperforms compared to GPT-4o in 24\% of cases. Additionally, o1-mini frequently fails to adhere to the expected output formats. This tendency contributes significantly to its limitations, with nearly 40\% of o1-mini's incorrect responses arising from improper output formats, an issue that is not observed with GPT-4o. Furthermore, when provided with human-written prompts, reasoning models like o1-mini may need additional time to interpret these instructions, potentially causing errors.}

\subsection{RQ3: the recommendations on how to select prompt engineering techniques and foundational LLMs}

In RQ1, we revisit the previous state-of-the-art prompt engineering techniques utilized in code-related tasks. The results show the effectiveness of each prompt engineering technique in different foundational LLMs (i.e., GPT-4o and o1-mini). In RQ2, we further investigate the advantages and limitations of the reasoning model in code-related tasks compared to non-reasoning LLMs. Hence, in RQ3, we want to offer practical guidance on how to select prompt engineering techniques and foundational LLMs when taking the monetary and time costs into account. 

\subsubsection{The cost of prompt engineering techniques when utilizing different foundational LLMs}

We first analyze the computational and token-based costs of different approaches for code generation, code translation, and code summarization, comparing them across two selected LLMs (i.e., GPT-4o and o1-mini). We focus on three key metrics: message tokens, reasoning tokens, and time cost (in seconds). Given the uncontrollable nature of reasoning time and CoT processes in reasoning LLMs, we measure the cost by calculating the call time and token usage of the response phase and reasoning phase separately. The evaluations are performed in a single-threaded environment on the same machine to ensure consistency.

\begin{table*}[t]
    \caption{Cost comparison of different approaches in code generation.}
    \label{tab:RQ3_gene}
    \resizebox{0.9\textwidth}{!}{%
    \begin{tabular}{lccc|ccc}
    \toprule
       & \multicolumn{3}{c|}{GPT-4o} & \multicolumn{3}{c}{o1-mini}  \\ \midrule
       & Message Token & Reasoning Token & Time Cost (s) & Message Token & Reasoning Token & Time Cost (s) \\
       \midrule
\multicolumn{1}{l|}{Zero-shot}  & 369.21 & 0 & 6.55 & 919.71  & 636.10 &  9.62 \\
\multicolumn{1}{l|}{CoT} & 400.75 & 0 & 15.83 & 1269.16 & 539.71 & 13.19   \\ 
\multicolumn{1}{l|}{AgentCoder} &  615.01 & 0 & 9.98 & 1747.54 & 1161.76 & 25.36  \\ 
\multicolumn{1}{l|}{Self-collaboration} & 732.85 & 0 & 13.91 & 1858.53 & 890.54 & 17.57 \\ 
\multicolumn{1}{l|}{LDB} & 702.87   & 0& 19.39 &  2625.74 & 1824.27 & 39.12 \\ 
\bottomrule
    \end{tabular}
    }
\end{table*}

\begin{table*}[t]
    \caption{Cost comparison of different approaches in code translation.}
    \label{tab:RQ3_trans}
    \resizebox{\textwidth}{!}{%
    \begin{tabular}{llccc|ccc}
    \toprule
  &  & \multicolumn{3}{c|}{GPT-4o} & \multicolumn{3}{c}{o1-mini}  \\ \midrule
    &   & Message Token & Reasoning Token & Time Cost (s) & Message Token & Reasoning Token & Time Cost (s) \\
       \midrule
 &\multicolumn{1}{l|}{zero-shot}& 108.00 & 0 & 4.35 & 122.35 & 401.93 & 6.49\\
& \multicolumn{1}{l|}{1-shot} & 98.14 & 0 & 3.61 & 111.78 & 631.59 & 9.05 \\
\multicolumn{1}{c}{Java2Python} & \multicolumn{1}{l|}{S\&G} & 259.04 & 0 & 9.32 & 294.94 & 669.38 & 8.31 \\
& \multicolumn{1}{l|}{Unitrans} & 572.64 & 0 & 15.12 & 484.84 & 8835.72 & 78.50 \\
& \multicolumn{1}{l|}{LostinTransl} & 639.79 & 0 & 18.53 & 673.85 & 5138.89 & 90.72  \\ \midrule
 &\multicolumn{1}{l|}{zero-shot}&  192.37 & 0 & 5.94 & 188.49 & 477.74 & 8.85  \\
& \multicolumn{1}{l|}{1-shot} & 123.35 & 0 & 3.63 & 140.93 & 672.00 & 9.88  \\
\multicolumn{1}{c}{Python2Java} & \multicolumn{1}{l|}{S\&G} & 346.52 & 0 & 7.73 & 418.05 & 708.38 & 10.80 \\
& \multicolumn{1}{l|}{Unitrans} & 694.85 & 0 & 12.09 & 994.69 & 9507.75 & 174.92  \\
& \multicolumn{1}{l|}{LostinTransl}  & 681.40 & 0 & 14.41 & 698.93 & 3829.95 & 74.53 \\ 

\bottomrule
    \end{tabular}
    }
\end{table*}

\begin{table*}[t]
    \caption{Cost comparison of different approaches in code summarization.}
    \label{tab:RQ3_summ}
    \resizebox{\textwidth}{!}{%
    \begin{tabular}{llccc|ccc}
    \toprule
  &  & \multicolumn{3}{c|}{GPT-4o} & \multicolumn{3}{c}{o1-mini}  \\ \midrule
    &   & Message Token & Reasoning Token & Time Cost (s) & Message Token & Reasoning Token & Time Cost (s) \\
       \midrule
& \multicolumn{1}{l|}{Zero-shot}  & 24.96 & 0 & 1.27 & 93.42 & 209.66 & 3.06  \\
& \multicolumn{1}{l|}{Few-shot}   & 19.65 & 0 & 1.33 & 51.72 & 302.59 & 3.96 \\
\multicolumn{1}{c}{Java} & \multicolumn{1}{l|}{CoT}  & 405.44 & 0 & 6.80 & 1041.77 & 582.91 & 11.41\\
& \multicolumn{1}{l|}{Critique}   & 193.05 & 0 & 5.63 & 393.36 & 1248.51 & 11.36 \\
& \multicolumn{1}{l|}{Expert} & 26.58 & 0 & 1.33 & 94.80 & 215.81 & 4.63  \\
& \multicolumn{1}{l|}{ASAP}   & 34.28 & 0 & 1.31 & 267.78 & 1204.48 & 8.65  \\ \midrule
& \multicolumn{1}{l|}{Zero-shot}  & 28.40 & 0 & 1.32 & 118.71 & 240.38 & 2.92 \\
& \multicolumn{1}{l|}{Few-shot}   & 16.56 & 0 & 1.38 & 44.69 & 344.58 & 10.81  \\
\multicolumn{1}{c}{Python} & \multicolumn{1}{l|}{CoT}  & 419.72 & 0 & 8.71 & 1166.48 & 646.66 & 13.23\\
& \multicolumn{1}{l|}{Critique}   & 213.94 & 0 & 8.11 & 449.15 & 1385.22 & 18.95  \\
& \multicolumn{1}{l|}{Expert} & 24.76 & 0 & 1.35 & 112.44 & 244.99 & 5.15 \\
& \multicolumn{1}{l|}{ASAP}   & 29.43 & 0 & 1.40 & 225.65 & 876.29 & 10.90  \\ 
 \bottomrule
    \end{tabular}
    }
\end{table*}

As shown in Table~\ref{tab:RQ3_gene}, \ref{tab:RQ3_trans}, and \ref{tab:RQ3_summ}, we can find when leveraging the same prompt engineering technique, \textbf{GPT-4o costs less tokens and time} compared to o1-mini. 
When utilizing the same foundational LLM, the kind of simple promptings, such as zero-shot and 1-shot promptings, cost the least among all the prompting engineering techniques. 
The prompting engineering techniques with multiple iterations need more computational and token-based costs compared to the prompting engineering techniques with single calling. 

\paragraph{Token usage} For o1-mini, the lengths of responses are usually longer than GPT-4o's responses even without considering reasoning tokens, as mentioned in RQ2. When considering reasoning tokens, the cost of o1-mini increases rapidly. For Unitrans and LostinTransl, which contain multiple iterations, the number of reasoning tokens is 5.5 times to 18.2 times compared to the number of message tokens. 
The fact that more complex prompt engineering techniques based on o1-mini tend to generate a larger volume of tokens suggests a correlation between prompt complexity and token generation. 
Furthermore, from the perspective of replies, this type of lengthy response also does not have significance in certain scenarios. For example, in code summarization, the average number of comment tokens for Java in CodeSearchNet~\cite{husain2019codesearchnet} is 36.54. 
However, the number of message tokens in o1-mini is much longer, often exceeding 1,000 when utilizing CoT prompting. This can lead to significant redundancies, which may interfere with processing and user understanding.

\paragraph{Time cost} When examining time cost, zero-shot prompting remains the most efficient, requiring the least amount of time. In comparison to approaches leveraging GPT-4o, approaches leveraging o1-mini result in substantially increased time costs. For instance, in code translation, LostinTransl's time cost with GPT-4o is 2.4 to 4.3 times higher than that of zero-shot prompting. 
However, when o1-mini is used, LostinTransl’s time cost escalates dramatically, ranging from 8.4 to 14.0 times that of zero-shot prompting. A similar trend is observed with Unitrans in Python-to-Java translation, where the time cost relative to zero-shot prompting increases from 2.0 to 19.8 times, without any corresponding improvement in performance. These observations suggest that advanced prompt engineering techniques may inadvertently extend the logical reasoning process of LLMs with built-in CoT, leading to longer reasoning times. 
This extended reasoning can result in significantly higher computational costs,
potentially outweighing the intended benefits of prompt complexity. 

It is also noteworthy that when considering the performance of GPT-4o and o1-mini (as detailed in RQ1), \textbf{the increased computational overhead—whether in token cost or time—does not necessarily translate into better performance}. This observation indicates that more complex prompts may inadvertently extend the reasoning process, leading to higher costs without corresponding benefits. For example, in the code summarization task, when using S\&G approach with GPT-4o and o1-mini, the performance actually decreases with the rise of costs.

\finding{GPT-4o is more efficient than o1-mini in both token usage and processing time. For example, in code translation, using LostinTransl increases GPT-4o's time cost by up to 4.3 times over zero-shot prompting, while for o1-mini, the time cost increases up to 14 times. Additionally, o1-mini's reasoning tokens can be up to 18 times the message tokens with complex techniques like Unitrans.
Additionally, this increased computational overhead does not necessarily translate into better performance, indicating that more complex prompts may inadvertently prolong reasoning, leading to greater costs without proportional benefits.
}

\subsubsection{The suggestions of selecting prompt engineering techniques and foundational LLMs}

We provide practical guidance for both practitioners and researchers to make informed decisions regarding the adoption of prompt engineering techniques and advanced LLMs when taking cost and effectiveness into account. 
In addition to computational and token-based costs, the carbon footprint of different prompt engineering techniques and foundational LLMs is a critical consideration. Each LLM call involves significant computational processes that consume electrical energy. Depending on the source of the electricity, this consumption can lead to varying levels of carbon emissions. Generally, higher token usage, longer reasoning times, and increased call duration translate to greater energy consumption~\cite{patterson2021carbon, xu2023energy, lazarev2022eco2ai}, which have significant environmental implications~\cite{llmcarbon2023, jagannadharao2023timeshifting}.

\paragraph{Foundational LLMs}
Based on the findings from RQ2, we suggest selecting LLMs based on the complexity of tasks. It can be preliminarily assessed by checking the length of CoT steps provided on the OpenAI website. To assess task complexity accurately, we recommend sampling a few representative data points directly from the task itself.

\textbf{For tasks that do not require complex reasoning} where the CoT length is less than 3 steps, \textbf{we recommend using GPT-4o} due to its relatively lower carbon footprint. 
Furthermore, in generating natural language for SE tasks, GPT-4o, combined with appropriate prompt techniques, can outperform reasoning LLMs with built-in CoT capabilities\footnote{A similar finding is observed in a previous NLP comparison~\cite{o1}.}. 
Additionally, users can select foundational LLMs based on the expected output length. For tasks with concise expected outputs, non-reasoning LLMs are more efficient choices.

In more \textbf{complex tasks} where the length of o1-mini CoT summaries is longer than or equal to 5 where o1-mini’s deeper reasoning capabilities are necessary, \textbf{we recommend utilizing LLMs with built-in CoT} (e.g., o1-mini). When utilizing LLMs with built-in CoT, we encourage the use of simple prompting techniques, such as zero-shot and one-shot, which can reduce energy consumption without substantially compromising performance. When the extra information can be obtained, utilizing LLMs with built-in CoT can further enhance the performance through a more comprehensive analysis than GPT-4o. 
Furthermore, when utilizing reasoning LLMs, users are recommended to write prompts with restrictions to ensure that reasoning LLMs maintain the original content of the input and avoid generating undesirable outputs.

\paragraph{Prompt Engineering Techniques}
For the advanced LLMs, the influence of the format of prompts is small. The prompting engineering \textbf{techniques including more information} tend to yield better performance. When generating code in SE tasks, the techniques that combine with the feedback of execution can more easily generate correct code. When generating texts in SE tasks, the prompt engineering techniques with iteration and more extracted information can achieve better performance based on GPT-4o. When generating texts in SE tasks, we do not recommend utilizing prompting engineering techniques on LLMs with built-in CoT at present (e.g., o1-mini).
When LLMs are not familiar with the format of output, an example to guide the format is necessary for prompt engineering techniques.

\finding{
For the selection of foundational LLMs in complex tasks requiring deep reasoning, the built-in CoT capabilities of reasoning LLMs can enhance performance. However, it is advisable to pair these capabilities with simpler prompting strategies, such as zero-shot or one-shot, to minimize carbon consumption. Regarding the selection of prompt engineering techniques, those that incorporate feedback or additional information can significantly improve code generation and text outputs. In other cases, we recommend opting for GPT-4o, since the prompt engineering techniques do not yield substantial benefits on reasoning LLMs.}

\section{Discussion}

In light of our findings on the reasoning capabilities of reasoning models, future work could explore several research directions to further harness and refine the strengths of reasoning LLMs in code-related and other complex tasks in SE. 

Optimizing prompt strategies for reasoning LLMs is an area ripe for exploration. Previous prompt engineering techniques designed for non-reasoning LLMs may not fully utilize the autonomous thinking and error-correction abilities of reasoning LLMs. Thus, devising adaptive prompt engineering techniques that integrate these capabilities could unlock new levels of performance. Researchers may experiment with minimalistic or constraint-based prompts that allow reasoning models to leverage their CoT while maintaining focus and avoiding extraneous reasoning steps.

Another promising avenue is the dynamic control of CoT length. While o1 can adjust its reasoning steps according to problem complexity, proper guidance could lead to more precise outputs that balance detail and efficiency. Researchers could investigate adaptive mechanisms that limit or expand CoT length based on predefined task characteristics, ensuring that reasoning LLMs do not overcomplicate simple tasks or underperform on challenging ones. By controlling the length and depth of reasoning, it may be possible to reduce computational costs while retaining high accuracy.

Additionally, ensuring o1’s outputs without unnecessary deviation or verbosity presents another opportunity for improvement in specific tasks. Techniques that align o1’s output to task-specific requirements could help reduce the model’s tendency toward excessive responses. This could be achieved by developing new prompt engineering techniques that guide the model’s output to remain concise and relevant, especially in tasks like code summarization~\cite{sun2024source, ahmad2023summarize} or commit message generation~\cite{tian2022makes, wanghard} where preciseness and conciseness are both critical.

Furthermore, it would be beneficial to explore methods for making reasoning models more cost-effective and environmentally sustainable. This could involve reducing token usage in CoT or implementing methods for batch-processing similar reasoning paths to minimize redundant computations. As reasoning models continue to evolve, considering the trade-offs between performance, cost, and environmental impact will be essential to their responsible deployment.

Overall, these future research directions, i.e., optimizing prompt strategies to enhance reasoning LLMs, controlling reasoning depth, and aligning output to task requirements, could lead to a more powerful and efficient use of reasoning LLMs in diverse SE applications.

\section{Threats to Validity}

\textbf{Threats to internal validity} mainly lie in implementing selected approaches of our experiments. To mitigate this threat, we directly use the released code of these approaches~\cite{sun2024source,zhong2024debug,huang2023agentcoder,dong2024self, yang2024exploring}. Another threat may appear due to the randomness of LLMs' response. To alleviate this threat, we run each approach three times and present the average results for each experiment. Additionally, the variations in the tuning parameters of these models may also affect their performance, which we controlled by adhering closely to the parameter settings recommended in their respective studies.

\textbf{Threats to external validity} lie in the dataset, which may influence the generalization of our findings. To mitigate this risk, we employ HumanEval~\cite{chen2021codex}, Transcoder~\cite{yang2024exploring}, and CodeSearchNet~\cite{husain2019codesearchnet,codexglue}, which are all widely used datasets in each task. However, due to the high cost of utilizing LLMs, our evaluation only involves three code-related tasks, i.e., code generation, translation, and summarization, and two programming languages, i.e., Java and Python. The generalization of our findings to other tasks and programming languages remains uncertain. To address this issue, preliminary tests in related tasks not detailed in this paper have been conducted, yielding similar results that support the conclusions drawn in this study. However, these results are not comprehensive enough to assert broad applicability. Future work will therefore involve more extensive evaluations across various tasks and programming languages to thoroughly assess the findings' generalizability.

\textbf{Threats to construct validity} lie in the metrics we used. In the three code-related tasks evaluated in our experiment, we utilize the most widely used metric in commit generation and code translation. In code summarization, common automated evaluation methods often lack consistency with human evaluations, which are mentioned in many code understanding tasks~\cite{wanghard, eliseeva2023commit, shi-etal-2022-race}. Hence, we follow the GPT-based evaluation methods proposed by Sun et al.~\cite{sun2024source}, which has demonstrated a stronger correlation with human judgments.

\section{Conclusion}

In conclusion, this study provides a comprehensive evaluation of prompt engineering techniques within the context of advanced large language models (including reasoning models and non-reasoning models) for SE tasks, focusing on code generation, translation, and summarization. Our results indicate that while prompt engineering has been essential in enhancing the effectiveness of earlier LLMs, its benefits are often diminished or altered when applied to more advanced models like GPT-4o and the reasoning LLM. Specifically, we find that reasoning models offer advantages primarily in complex tasks requiring multi-step reasoning but may not justify their additional costs and potential environmental impact in simpler tasks where non-reasoning models perform comparably or even more effectively.
Our findings suggest that adapting prompt engineering techniques to advanced LLMs requires a shift in focus, emphasizing accurate information input and response evaluation rather than complex prompt structures. For SE tasks that do not heavily rely on reasoning, simpler prompt configurations with non-reasoning models can deliver high-quality results with greater cost efficiency. Additionally, when using reasoning models for more intricate tasks, careful management of output format and content length is advised to improve usability and relevance.
This study contributes valuable insights into the evolving landscape of LLM applications in SE, underscoring the importance of adapting prompt engineering strategies in line with the capabilities and limitations of current LLM advancements. Future research may further explore the nuanced interplay between prompt complexity and model capabilities, providing deeper insights into optimizing LLM deployment across a broader array of SE applications.

% \section{Acknowledgments}

\section{Data Availability}

All data and results are available on our homepage~\cite{anonymous_2024_14023397}.

\bibliographystyle{ACM-Reference-Format}
\bibliography{sample-base}

\end{document}